\documentclass[epj]{svjour}
\usepackage{graphicx}
\begin{document}

\title{Scaling of broadband dielectric data of glass-forming liquids and plastic
crystals}
 \titlerunning{Scaling of broadband dielectric data...}
 \author{U. Schneider, R. Brand, P. Lunkenheimer, and A. Loidl}
 \authorrunning{U. Schneider {\it et al.}}
\institute{Experimentalphysik V, Universit\"{a}t Augsburg, D-86135
Augsburg, Germany}
\date{submitted to European Physical Journal E}
%
\abstract{The Nagel-scaling and the modified scaling procedure proposed recently by
Dendzik \textit{et al.} have been applied to broadband dielectric data on two
glass-forming liquids (glycerol and propylene carbonate) and three plastic crystals
(ortho-carborane, meta-carborane, and 1-cyano-adamantane). Our data extend the upper
limit of the abscissa range to considerably higher values than in previously published
analyses. At the highest frequencies investigated, deviations from a single master curve
show up which are most pronounced in the Dendzik-scaling plot. The loss curves of the
plastic crystals do not scale in the Nagel-plot, but they fall onto a separate master
curve in the Dendzik-plot. In addition, we address the question of a possible
divergence of the static susceptibility near the Vogel-Fulcher temperature.
For this purpose, the low-temperature evolution of the high-frequency wing of the
dielectric loss peaks is investigated in detail. No convincing proof for such a
divergence can be deduced from the present broadband data.
} 
\maketitle

\section{Introduction}

Investigations of the dynamic response of glass-forming materials are of fundamental importance to describe glass
transition phenomena \cite{ngai94}. Here the $\alpha $- or structural relaxation is the most commonly investigated
dynamic process. In the frequency dependence of the dielectric loss, $\varepsilon ^{\prime \prime }(\nu )$, of dipolar
systems the $\alpha $-peak is the dominating spectral response, strongly slowing down with decreasing temperature. The
microscopic origin of the spectral shape of the $\alpha $-peak, which significantly deviates from the mono-dispersive
Debye behaviour, is still unclear in many respects. Usually phenomenological functions as, e.g., the Cole-Davidson (CD)
function \cite{davidson51} or the Fourier transform of the Kohlrausch-Williams-Watts (KWW) function \cite
{kohlrausch54} are used to describe the $\alpha $-peak. However, these empirical functions, which extrapolate to a
power law, $\varepsilon ^{\prime \prime }\sim \nu ^{-\beta }$, $\beta <1$, at high frequencies, provide good fits of
the experimental data at best up to 3 decades above the peak frequency, only. At higher frequencies an excess wing
appears, which can be described by a second power law, $\varepsilon ^{\prime \prime }\sim \nu ^{-b} $ with $b<\beta $
\cite{hofmann94,lunkenheimer96,leheny97,leheny98}. This excess wing was already noted in the early work of Davidson and
Cole \cite{davidson51} and is a universal feature of glass-forming liquids
\cite{hofmann94,lunkenheimer96,leheny97,leheny98,dixon90}, at least for all systems where a $\beta $-relaxation is
absent (see below). Recently, we were able to show that this excess wing does not show up in various plastic crystals
\cite{brand99}, a class of materials, whose properties resemble those of glass-forming liquids in many respects. In
plastic crystals the centres of mass of the molecules form a regular crystalline lattice but the molecules are
dynamically disordered with respect to the orientational degrees of freedom. Due to similarities in the relaxation
dynamics, orientationally disordered crystals are often considered as model systems for structural glasses, but are
much simpler to be treated in theoretical approaches.

It has to be stated that until now there is no commonly accepted explanation of the microscopic origin of the excess
wing in glass-forming liquids, although there are some theoretical approaches that are able to describe the wing at
least partly \cite{chamberlin93}. In addition, it is possible to describe $\alpha $-peak and wing by a superposition of
two relaxation processes \cite{hofmann94,Leon99a,Leon99b,Lunkihab}. A secondary relaxation process, usually termed $
\beta $-process, was demonstrated to be present in a variety of glass formers by Johari and Goldstein \cite{Johari70}.
Also recent theoretical developments within the coupling model \cite{ngai79} may point in the direction of a universal
$\beta$-relaxation, closely connected to the $\alpha $-process \cite{ngai98,Leon99b} and it cannot be excluded that the
wing is caused by a $\beta $-relaxation, whose peak is hidden under the dominating $\alpha $-peak. Of course, as long
as the occurrence of the excess wing and its evolution with temperature remains a mystery, also its absence in plastic
crystals cannot be explained.

A very successful description of the wing is given by the so-called Nagel-scaling
\cite{dixon90,menon92}. Nagel and coworkers found that $\varepsilon ^{\prime \prime }(\nu
)$-curves for different temperatures and even for different materials, including the
wing, can be scaled onto one master curve by plotting $Y_{N}:=1/w\log _{10}[\varepsilon
^{\prime \prime }\nu _{p}/(\Delta \varepsilon \,\nu )]$ \textit{vs.}
$X_{N}:=1/w(1+1/w)\log _{10}(\nu /\nu _{p})$. Here $w$ denotes the half-width of the loss
peak normalized to that of the Debye-peak, $\nu _{p}$ is the peak frequency, and $\Delta
\varepsilon $ the relaxation strength. During the past years some criticism of the
Nagel-scaling arose concerning its universality \cite {schoenhals91,dendzik97} and
accuracy \cite{dendzik97,kudlik95}. However, it is still commonly believed that the
Nagel-scaling is of significance for our understanding of glass-forming liquids and many
efforts have been made to check for its validity in a variety of materials \cite
{brand99,menon92,schoenhals91,dendzik97,kudlik95,leslie94,menon93,leheny96}. Recently a
modification of the original scaling procedure was proposed by Dendzik \textit{et al.}
\cite{dendzik97} and reported to lead to a better scaling, especially at low frequencies.
Here $Y_{D}:=\log _{10}[\varepsilon^{\prime \prime }\nu _{s}/(\varepsilon _{s}\nu )]$ is
plotted \textit{vs.} $X_{D}:=(1+\beta )\log _{10}(\nu /\nu _{s})$ \noindent with
$\nu_{s}$ and $\varepsilon _{s}$ read off at the intersection point of the two power laws
$\varepsilon ^{\prime \prime }\sim \nu $ and $\varepsilon ^{\prime \prime }\sim \nu
^{-\beta }$ below and above $\nu _{p}$, respectively.

The universal scaling of $\alpha $-peak and excess wing strong\-ly suggests a correlation of both features. Especially,
making certain assumptions, it was shown \cite{menon95}, that the\ wing exponent\ $b$ should become zero for a limiting
relative half width, $w^{\ast }\approx 2.6$. It was argued, that this constant loss behaviour could imply a divergence
of the static susceptibility, supporting speculations about an underlying phase transition from the liquid to the
glassy state. By a low temperature extrapolation of $w(T)$ curves of various glass-formers, Nagel and coworkers
deduced, that this constant-loss limit is reached near the Vogel-Fulcher temperature $T_{VF}$ \cite{menon95}. $T_{VF}$
is determined from fits of the $\alpha $-relaxation time $\tau $ with the Vogel-Fulcher law, $\tau =\tau _{0}\exp
[DT_{VF}/(T-T_{VF})]$. In addition, it was shown \cite{leheny97} that, if the scaling is valid and assuming $w=1/\beta
$, the exponents $\beta $ and $b$ are related by
\begin{equation}
\frac{b+1}{\beta +1}=\gamma ,  \label{gamma}
\end{equation}
where $\gamma $ should be equal to the high-frequency limiting slope in the Nagel-plot, $\gamma^\prime$. Indeed, in
\cite{leheny97} within the error bars a temperature-independent value $\gamma =0.72(\pm 0.02)$ was found, in agreement
with $\gamma^\prime$ read off from the master curve of the Nagel-scaling. Finally, a ''divergent'' susceptibility was
also deduced directly from an ex\-tra\-po\-la\-tion of $b$ to low temperatures, which for various glass formers indeed
seemed to approach zero near $T_{VF}$ \cite{leheny97}.

Our group has made available dielectric loss data on glass-forming liquids and plastic crystals extending over the
exceptionally broad dynamic range of up to 19 frequency decades
\cite{lunkenheimer96,lunkenheimer97,lunkenheimer96b,lunkenheimer97b,schneider98,schneider99}. The present work has two
objectives: First, to check if scaling still works for these extremely broadband data. Second, to look for evidence for
a divergent dielectric susceptibility at low temperatures. We will apply both, the original Nagel-scaling \cite
{dixon90} and the modified procedure proposed by Dendzik \textit{et al.} \cite{dendzik97} to broadband dielectric data
on the low-molecular weight glass-formers glycerol and propylene-carbonate (PC) and on the plastic crystals
ortho-carborane (o-CA), meta-carborane (m-CA), and 1-cyano-adamantane (CNA). In addition, these data will be checked
for a cross-over of the high-frequency wing to constant loss behaviour at low temperatures.

\section{Experimental}

To collect data in a broad frequency range, various experimental setups were used, including a time domain spectrometer
($10\,\mu \mathrm{Hz}\,\leq \nu \leq \,1\,\mathrm{kHz}$), frequency response analyzers and LCR-meters
($10\,\mathrm{mHz}\,\leq \nu \leq \,30\,\mathrm{MHz}$), impedance analyzers (reflectometric technique,
$1\,\mathrm{MHz}\,\leq \nu \leq \,1.8\,\mathrm{GHz}$), a network analyzer (reflection and transmission,
$100\,\mathrm{MHz}\,\leq \nu \leq \,30\,\mathrm{GHz}$), a quasi-optic submillimeter wavelength spectrometer
(transmission and phase measurement, $40\,\mathrm{GHz}\,\leq \nu \leq \,1.2\,\mathrm{THz}$), and a Fourier-transform
infrared spectrometer (transmission, $600\,\mathrm{GHz}\,\leq \nu \leq \,3\,\mathrm{THz}$). For more experimental
details the reader is referred to \cite{schneiderlees}.

\section{Results and discussion}

\begin{figure}[tbp]
\centering
 \resizebox{0.48\textwidth}{!}{\includegraphics[clip]{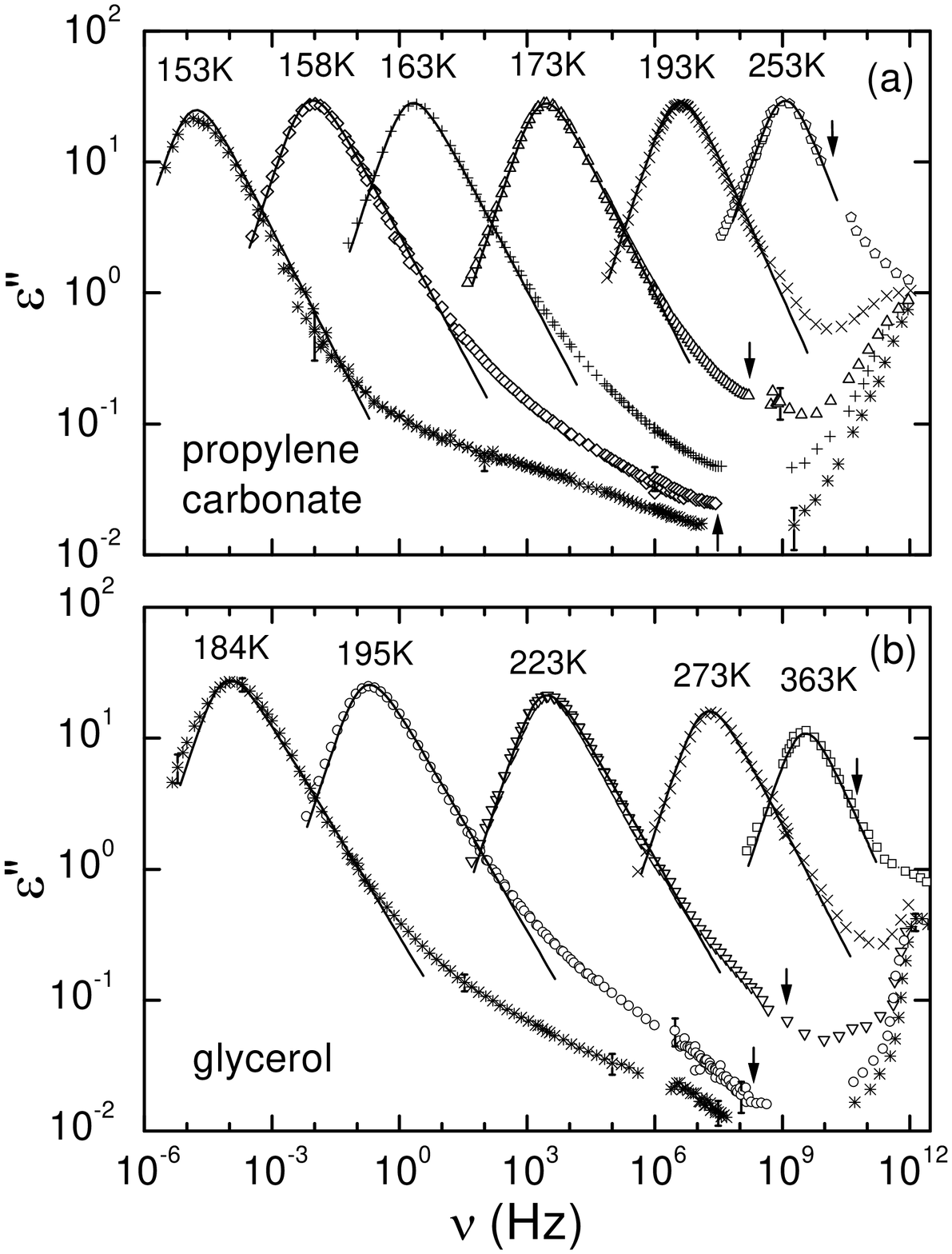}}
\caption{Frequency dependence of the dielectric loss of PC (a) and glycerol (b) for various temperatures. For clarity
reasons, only selected temperatures are shown and for frequencies $\nu > 1\,\mathrm{GHz}$ only a part of the recorded
data points are plotted (for the complete curves, see \protect\cite {schneider98,schneider99}). The solid lines are
fits with the CD function. The arrows indicate the cutoff frequency up to which the data have been used for the scaling
plots of Figure~\ref{sgf_dn}.} \label{sgf_e2}
\end{figure}

Figure~\ref{sgf_e2} shows the frequency dependence of the dielectric loss of PC and glycerol for various temperatures
\cite{schneider98,schneider99}.
$\varepsilon ^{\prime \prime }(\nu )$ exhibits the typical asymmetrically shaped $\alpha $-relaxation peaks shifting
through the frequency window with temperature. At high frequencies, in the GHz -- THz range, a minimum shows up,
followed by another loss peak (the microscopic or boson peak). These high-frequency features are treated in detail in
\cite {lunkenheimer97,lunkenheimer96b,lunkenheimer97b,schneider98,schneider99}. The solid lines in Figure~\ref{sgf_e2}
are fits of the $\alpha $-peak region with the empirical CD function \cite{davidson51}, which leads to good fits of the
peak region. It was shown before \cite {lunkenheimer97,lunkenheimer96b,lunkenheimer97b,schneider98,schneider99}, that
the CD function describes the data clearly better than the Fourier transform of the KWW function. At frequencies about
$2-3$ decades above $\nu _{p}$ the excess wing shows up as a deviation of $\varepsilon ^{\prime \prime }(\nu )$ from
the CD-fits, forming a second power law, $\varepsilon ^{\prime \prime }\sim \nu ^{-b}$. The slope of the excess wing
increases with increasing temperature as found previously for other glass-forming materials \cite{leheny97}. For high
temperatures the excess wing seems to merge with the $\alpha $-peak. At the lowest temperatures investigated, for both
materials in the excess wing region a slight downward curvature of $\varepsilon ^{\prime \prime }(\nu )$ may be
suspected. This could be indicative for a $\beta $-relaxation process. The associated $\beta $-peak could be either
superimposed to the excess wing, becoming important at low temperatures only \cite{paluch}, or it may even be the
origin of the whole wing-feature \cite{hofmann94,Leon99a,Leon99b,Lunkihab}. However, the experimental errors are too
large to allow for a definite conclusion concerning the presence of a $\beta $-peak and further experiments are in
progress to clarify this question.

\begin{figure}[tbp]
 \centering
\resizebox{0.48\textwidth}{!}{\includegraphics[clip]{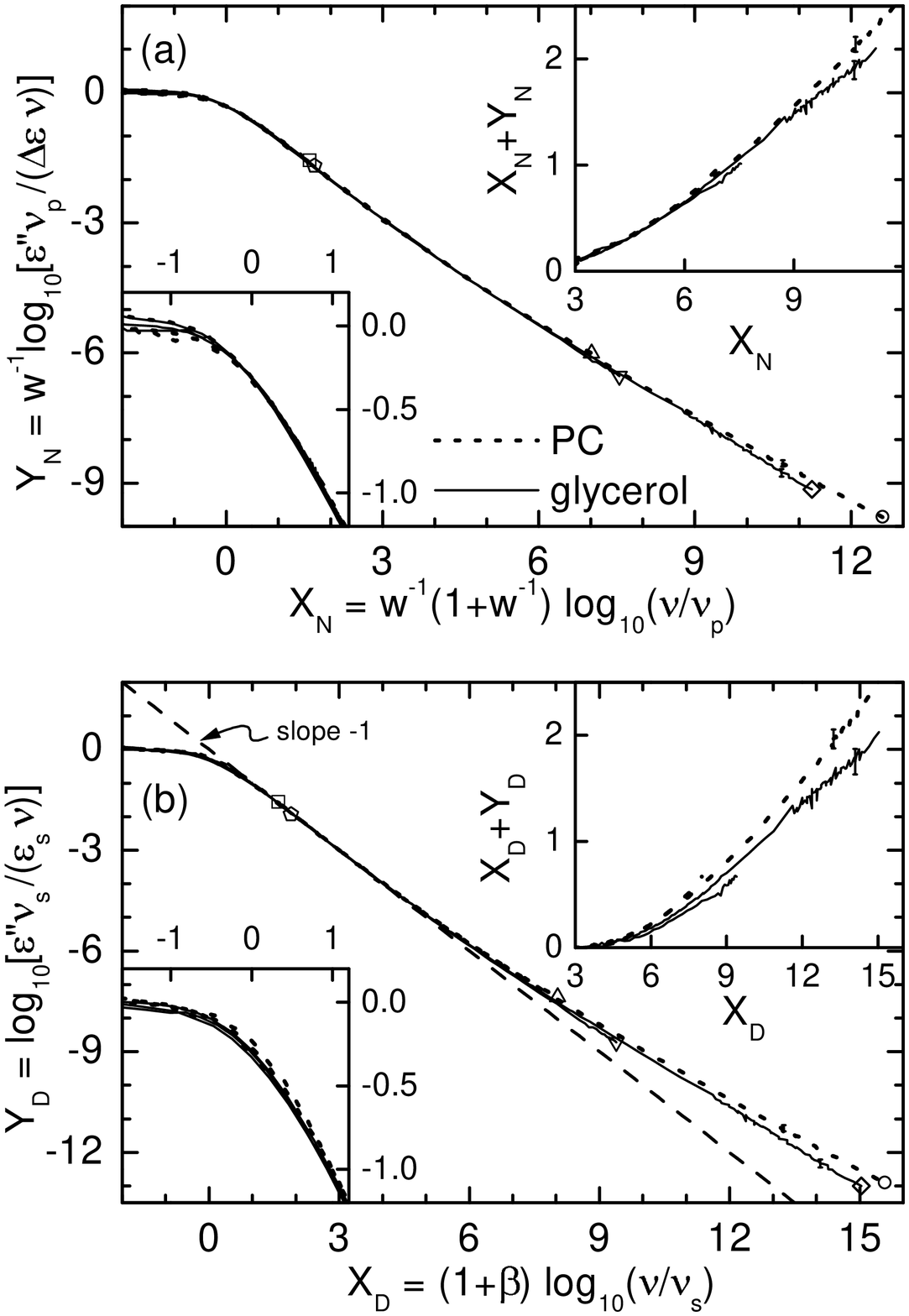}} \caption{(a) Nagel-plot \protect\cite{dixon90} of the
dielectric loss results on PC at 158 K, 173 K, and 253 K and glycerol at 195 K, 223 K, and 363 K. The lower inset shows
a magnified view of the region near $X=0$. The upper inset shows the data for $X>3$, with the ordinate replaced by the
sum of ordinate and abscissa of the main frame. The symbols (same as in Figure~\ref{sgf_e2}) indicate the highest value
of $X$ used for the different temperatures. (b) Same results as in (a), scaled according to Dendzik \textit{et al.}
\protect\cite{dendzik97}. The dashed line indicates the scaling curve for CD behavior at $\protect\nu
>\protect\nu_{s}$. The insets are defined in the same way as for (a).} \label{sgf_dn}
\end{figure}

The data on glycerol and PC, after application of the Nagel \cite{dixon90} and Dendzik-scaling \cite{dendzik97} are
shown in Figures~\ref{sgf_dn}(a) and (b), respectively.
To maintain readability, for each material $\varepsilon ^{\prime \prime }(\nu )$ curves for three different
temperatures are shown only. For both materials the lowest temperature has been omitted, as here, due to the relatively
high experimental errors near the $\alpha $-peak region, an unequivocal determination of the scaling parameters is
difficult. We also want to avoid contributions from a possible $\beta $-peak, superimposed to the wing at low
temperatures\footnote[1]{Indeed, for PC, even after an optimization of the Nagel-scaling parameters \cite{leheny96}
within the experimental errors, the 153 K curve shows pronounced deviations from the curves at higher temperatures in
the whole frequency range (not shown).} as, e.g., found for Salol \cite{kudlik95}. The data have not been corrected for
conductivity contributions, which only weakly affect $\varepsilon ^{\prime \prime }(\nu )$ about 1--2 frequency decades
below $\nu _{p}$. The effect of the conductivity and the correct method of subtraction are a matter of controversy
\cite{menon93,leheny96} and therefore we decided to show the raw data. The scaling cannot be expected to work in the
region of the $\varepsilon ^{\prime \prime }$-minimum in the GHz -- THz region. Therefore only data-points up to a
cutoff frequency that is indicated by arrows in Figure~\ref{sgf_e2} were used. The maximum value of $X$ used at the
different temperatures is indicated by the symbols. For the Dendzik-scaling, the present plots extend to higher values
of $X_{D}$ than any previously published scaling plots. For the Nagel-scaling, a plot up to $X_{N}=12$ has been
published in \cite{menon95} but for one material only (glycerol). In addition, a Nagel-plot up to $X_{N}=16$, comparing
different glass formers, was shown in \cite{leheny98}. However, this plot was obtained using data where the $\alpha
$-peak was missing and some crude assumptions for the determination of the scaling parameters had to be made to arrive
at the scaling at high frequencies. Figure~\ref{sgf_dn} reveals, that both scaling procedures lead to good scaling of
the data at not too high values of $X$. There are reports of some shortcomings of the Nagel-scaling in the region near
$X=0$ \cite{schoenhals91,dendzik97,kudlik95}. In contrast, the modified procedure proposed by Dendzik \textit{et al.}
was shown to lead to a better scaling in this region \cite{dendzik97}. For the present data on PC and glycerol in this
region [lower insets in Figures~\ref{sgf_dn}(a) and (b)] indeed the quality of the Nagel-scaling seems to be slightly
worse, but the differences are very subtle. It was shown that the modified procedure proposed by Dendzik \textit{et
al.} is able to scale different CD curves onto one master curve \cite{dendzik97}. At high frequencies this master curve
is a straight line with slope $-1$ [straight dashed line in Figure~\ref{sgf_dn}(b)]. Then the excess wing should show
up as deviations from the CD master curve at high $X_{D}$ values, as is indeed seen in Figure~\ref{sgf_dn}(b). The most
interesting result of these scaling plots is, that at high values of $X$, in both scaling approaches deviations from a
single master curve show up. These deviations are more pronounced in the Dendzik-scaling. For example, in the
Dendzik-scaling the curves at 158 K for PC and at 195 K for glycerol start to significantly deviate from each other
above an $X_{D}$-value corresponding to 50 Hz for PC and 1 kHz for glycerol. In the Nagel-plot this deviation occurs at
43 kHz for PC and at 3 MHz for glycerol. While for the Nagel-scaling the deviations are almost within the experimental
errors, this is not the case for the Dendzik-scaling (see error bars in Figure~\ref{sgf_dn}), which clearly fails at
high frequencies. However, for both procedures the curves obtained at different temperatures for a single material
scale (also including data at temperatures not shown in Figure~\ref{sgf_dn}).

A general problem of the scaling plots is the fact that the $Y$-value varies considerably, roughly following $Y=-X$,
but the information of interest is the presence of small deviations from this variation. Therefore a much better
resolution can be reached when correcting for this $Y=-X$ behaviour. This can be achieved by plotting the quantity
$Z(X)=Y(X)+X$. In the upper insets of Figures~\ref{sgf_dn}(a) and (b) the data at $X>3$ are plotted in this way.
Clearly the resolution of the plots is enhanced and the deviations of the different curves from a universal scaling
behaviour become more obvious.

Finally it may be mentioned that for high temperatures ($T \geq 253 \, $ K in glycerol and $T \geq 203 \, $ K in PC),
the dielectric loss data can also be scaled onto each other by simply plotting $\varepsilon/\varepsilon_p$ {\it vs.}
$\nu / \nu _p$ with $\varepsilon_p$ being the peak value of $\varepsilon^{\prime\prime}(\nu)$. Here the excess wing
seems to have merged with the $\alpha $-peak and $\beta $ is nearly temperature independent.

\begin{figure}[tbp]
\centering
 \resizebox{0.48\textwidth}{!}{\includegraphics[clip]{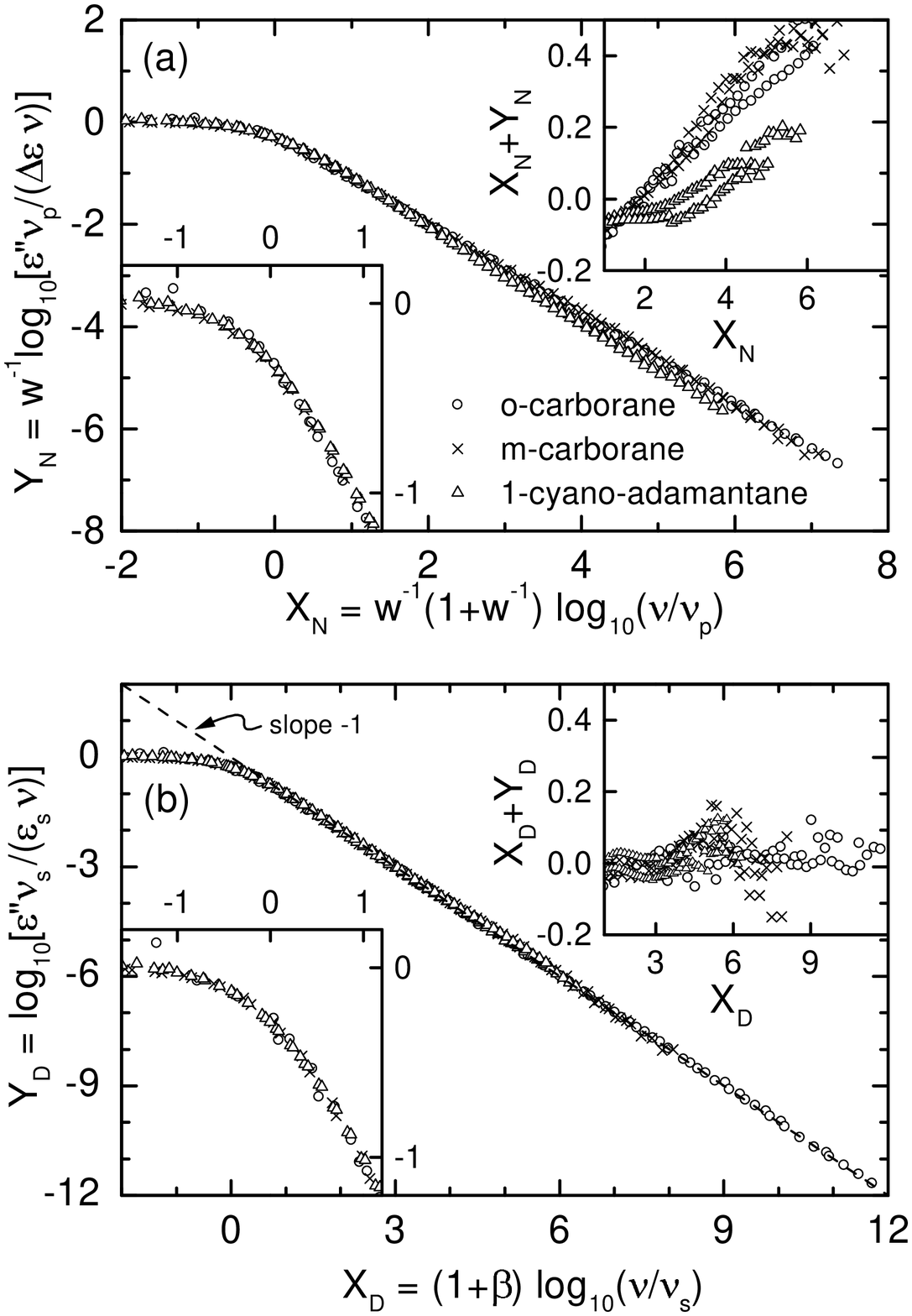}}
\caption{Same plot as in Figure~\ref{sgf_dn}, but for the plastic crystals ortho-carborane at 130 K and 150 K,
meta-carborane at 200 K and 210 K, and 1-cyano-adamantane at 280 K and 300 K \protect\cite{brand99}.} \label{px_dn}
\end{figure}

In Figure~\ref{px_dn} both scaling plots are shown for the plastic crystals (o-CA, m-CA, and CNA).
These plastic crystals have a rigid molecular structure and $\beta $-relaxation contributions from intramolecular
degrees of freedom can be excluded. Their dielectric loss was shown to exhibit no or only a very weak wing
\cite{brand99}. In Figure~\ref{px_dn}, for clarity reasons, curves for two temperatures only are shown for each
material. Clearly, the Nagel-scaling [Figure~\ref{px_dn}(a)] does not work for the plastic crystals investigated
\cite{brand99}. Again, in the upper inset of Figure~\ref{px_dn}(a) the deviations become more obvious. The deviations
occur at much lower $X_{N}$-values and are much stronger than for the glass-forming liquids [compare
Figure~\ref{sgf_dn}(a)]. This can be rationalized considering that the $\varepsilon ^{\prime \prime }(\nu )$-curves of
the plastic crystals all follow a CD behaviour quite closely up to the highest frequencies \cite{brand99} and that
different CD-curves do not scale in the Nagel-plot \cite{menon92}. In contrast, in the Dendzik-plot
[Figure~\ref{px_dn}(b)] all curves for the plastic crystals scale onto the CD master curve (dashed line). It should be
noted, that these plastic crystals are relatively ''strong'' glass-formers \cite{lunkenheimer96b,boehmer93,Brandunpubl}
within the classification scheme proposed by Angell and coworkers \cite{strong}. It cannot be excluded that the failing
of the Nagel-scaling in these materials is due to their high strength and not due to their plastic crystalline nature.
This would also be in accord with Ngai's ideas \cite{ngai98,Leon99b}, concerning an explanation of the excess wing
within the coupling model. Unfortunately, to our knowledge the Nagel-scaling was never applied to strong structural
glass formers.

\begin{figure}[tbp]
\centering
 \resizebox{0.4\textwidth}{!}{\includegraphics[clip]{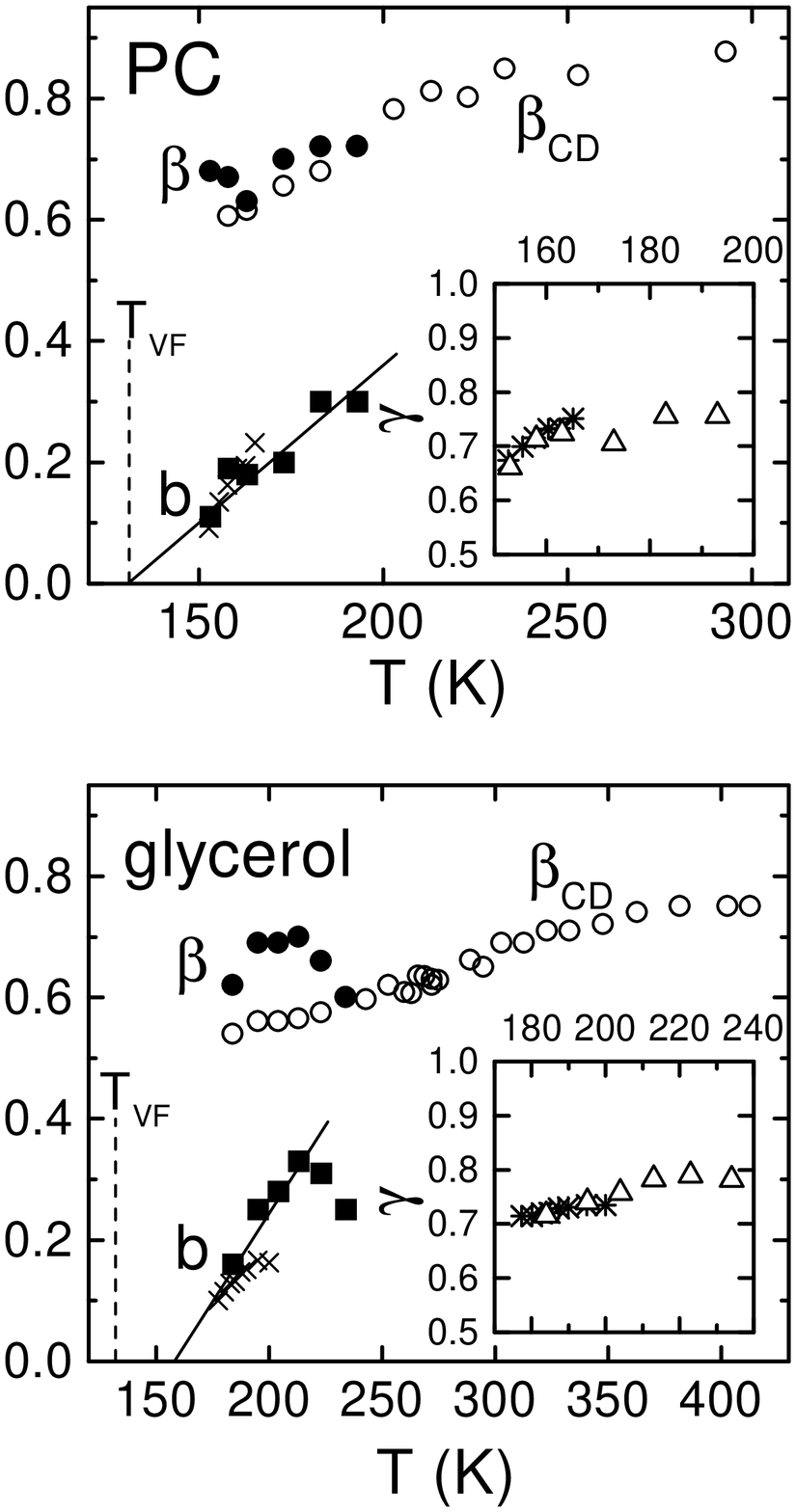}}
\caption{Temperature dependence of $\protect\beta _{CD}$ (open circles) from fits of the
$\protect\alpha $-peak with the CD function and of $\protect\beta $ (closed circles) and $b$
(squares) from fits with a phenomenological ansatz (see text) for PC and glycerol. The
crosses show the results reported in \protect\cite{leheny97}.
The straight solid lines demonstrate possible low-temperature extrapolations of $b(T)$.
$T_{VF}$ is indicated by the dashed line. The insets show $\protect\gamma $ as defined in
equation~(\ref{gamma}) (triangles). The results from \protect\cite{leheny97} are shown as
stars.} \label{beta}
\end{figure}

Finally, we want to check, if the excess wing approaches constant-loss behaviour at low temperatures in the samples
investigated. Figure~\ref{beta} shows the temperature dependence of $\beta _{CD}$ for PC and glycerol, as resulting
from the fits of the $\alpha $-peak with the CD function (Figure~\ref{sgf_e2}).
For the CD function, the limiting half width, $w^{\ast }\approx 2.6$, at which $\varepsilon''=$\,const \cite{menon95}
corresponds to $\beta ^{\ast }\approx 0.18$. Inspecting Figure~\ref{beta}, it seems unlikely that $\beta _{CD}(T)$ will
reach $\beta ^{\ast }$ at $T\approx T_{VF}$. From equation~(\ref{gamma}) and assuming $\gamma =\gamma^\prime =0.72$
\cite{leheny97}, a different limiting value for the divergence of the static susceptibility, $\beta ^{\ast \prime
}\approx 0.38$, can be deduced \cite {leheny97}. While a linear extrapolation of $\beta (T)$ leads to a somewhat higher
value of $\beta (T_{VF})$, an extrapolation to $\beta (T_{VF})=0.38$ at least is consistent with the data. The
discrepancy between $\beta ^{\ast } $ and $\beta ^{\ast \prime }$ arises due to the poor fulfillment of the assumption
$\beta =1/w$ made for the deduction of equation~(\ref{gamma}) from the Nagel-scaling. In contrast to this, for the
Dendzik-scaling, equation~(\ref{gamma}) should be valid without such an assumption. However, while for the
Nagel-scaling a sample-independent value of $\gamma^\prime =0.72$ is consistent with our data [Figure~\ref{sgf_dn}(a)],
for the Dendzik-scaling the value of $\gamma^\prime $, read off from the high-frequency slope in the scaling plot
becomes sample dependent. From Figure~\ref{sgf_dn}(b), $\gamma^\prime =0.79$ and $\gamma^\prime =0.715$ are determined.
If $\gamma = \gamma \prime$ is assumed still to be valid, Equation~\ref{gamma} leads to $\beta ^{\ast \prime }\approx
0.27$ and $\beta ^{\ast \prime }\approx 0.4$ for glycerol and PC, respectively. The experimental data of $\beta
_{CD}(T)$ may be consistent with these values (Figure~\ref{beta}).

A more direct way to check for a divergent static susceptibility is the investigation of the temperature dependence of
the exponent $b$. However, an unambiguous determination of $b$ is difficult, especially at higher temperatures, where
$\alpha $-peak, wing, and the minimum are strongly superimposed to each other. For this purpose, a parameterization
using the sum of two power laws was used in \cite{leheny97}. In order to take into account the additional contributions
in the minimum region and the increase towards the boson peak, a constant loss and two power laws have to be added
\cite{lunki99} leading to $\varepsilon''=c_{\beta}\nu^{-\beta}+c_{b}\nu^{-b}+\varepsilon_{c}+c_{3}\nu^{0.3}+c_{n}\nu
^{n}$ . This phenomenological ansatz was shown to provide good fits over a considerable frequency range \cite{lunki99}.
In contrast to the scaling plots, for this analysis the lowest-temperature curves have been included as at least the
$\nu^{-\beta}$ power-law  seems to be well defined. In addition, a possible $\beta $-peak, superimposed to the wing,
may be expected to have only a small influence on the exponent $b$ determined from the fit. The resulting $b(T)$ and
$\beta (T)$ are shown in Figure~\ref{beta}. $\beta (T)$ tends to assume somewhat higher values than $\beta _{CD}$ due
to the contributions of the wing power law $\nu ^{-b}$ at low frequencies. In Figure~\ref{beta} also the results of
Leheny and Nagel \cite{leheny97} are included (crosses). Due to the larger frequency range available to us and the
different evaluation procedure we were able to extend the results of these authors to considerably higher temperatures.
However, results of $b(T)$ could only be obtained up to 193 K for PC and up to 234 K for glycerol. Above these
temperatures the excess wing has almost merged with the $\alpha $-peak and an unequivocal determination of $b$ is no
longer possible.

In the insets of Figure~\ref{beta} the obtained results for $b$ and $\beta $\ are used to check the prediction $\gamma
=\gamma^\prime$ arising from the Nagel-scaling. For both materials $\gamma $\ agrees roughly with the value of
$\gamma^\prime =0.72 (\pm 0.02)$, found from the high-frequency limiting slope of the Nagel-scaling master curve
\cite{leheny97}, however, exhibiting a tendency to increase with temperature. Again, for comparison the results of
Leheny and Nagel \cite{leheny97} are included (stars). It is interesting that the slight temperature variation of
$\gamma $, seen in the data of Leheny and Nagel \cite{leheny97} agrees with that detected in the present results. This
indicates a systematic nature of this temperature dependence in contrast to the assumption of a constant $\gamma (T)$
within the error bars, made in \cite{leheny97}. The significant increase of $\gamma (T)$, detected for both materials,
seems reasonable in light of the merging of excess wing and $\alpha $-peak at high temperatures which can be suspected
from Figure~\ref{sgf_e2}. This implies the approach of $\gamma =1$ for high temperatures.

\begin{figure}[tbp]
\centering
 \resizebox{0.4\textwidth}{!}{\includegraphics[clip]{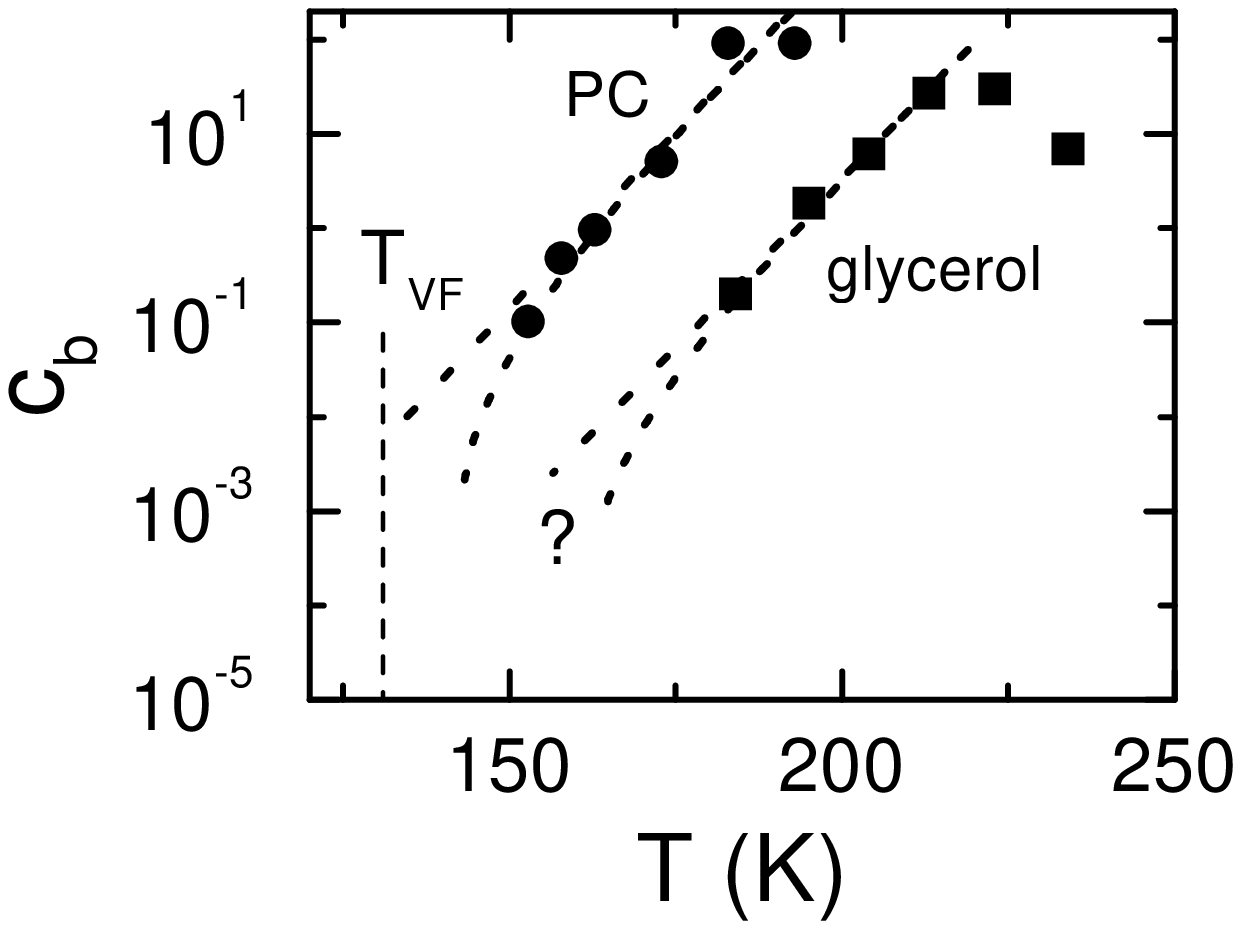}}
\caption{Prefactor $c_{b}$ of the power law  $c_{b}\protect\nu
^{-b}$ used for the description of the excess wing in glycerol (squares) and
PC (circles). The dashed line indicates $T_{VF}=131 $ K for glycerol
\protect\cite{schneider98}. For PC, $T_{VF}=132$ K was determined
\protect\cite{schneider99}. The dotted lines demonstrate that it is not possible to draw
conclusions concerning a non-zero value of $c_{b}$ at $T_{VF}$. } \label{cb}
\end{figure}

As demonstrated by the solid lines in Figure~\ref{beta}, $b(T)$ can be\ reasonably extrapolated to zero at $T_{VF}$ for
PC and to about $T_{VF}+25$ K for glycerol. The result for PC indeed seems to support the approach of a constant loss
at $T_{VF}$. However, even if $b$ becomes zero, a ''divergent'' susceptibility will only be reached, if the pre-factor
$c_{b}(T)$ does not approach zero for $T\rightarrow T_{VF}$. In Figure~\ref{cb} $c_{b}(T)$ is shown for both materials.
As indicated by the dashed lines the results do not allow for a definite conclusion concerning a finite value at
$T_{VF}$.

\begin{figure}[tbp]
\centering
 \resizebox{0.4\textwidth}{!}{\includegraphics[clip]{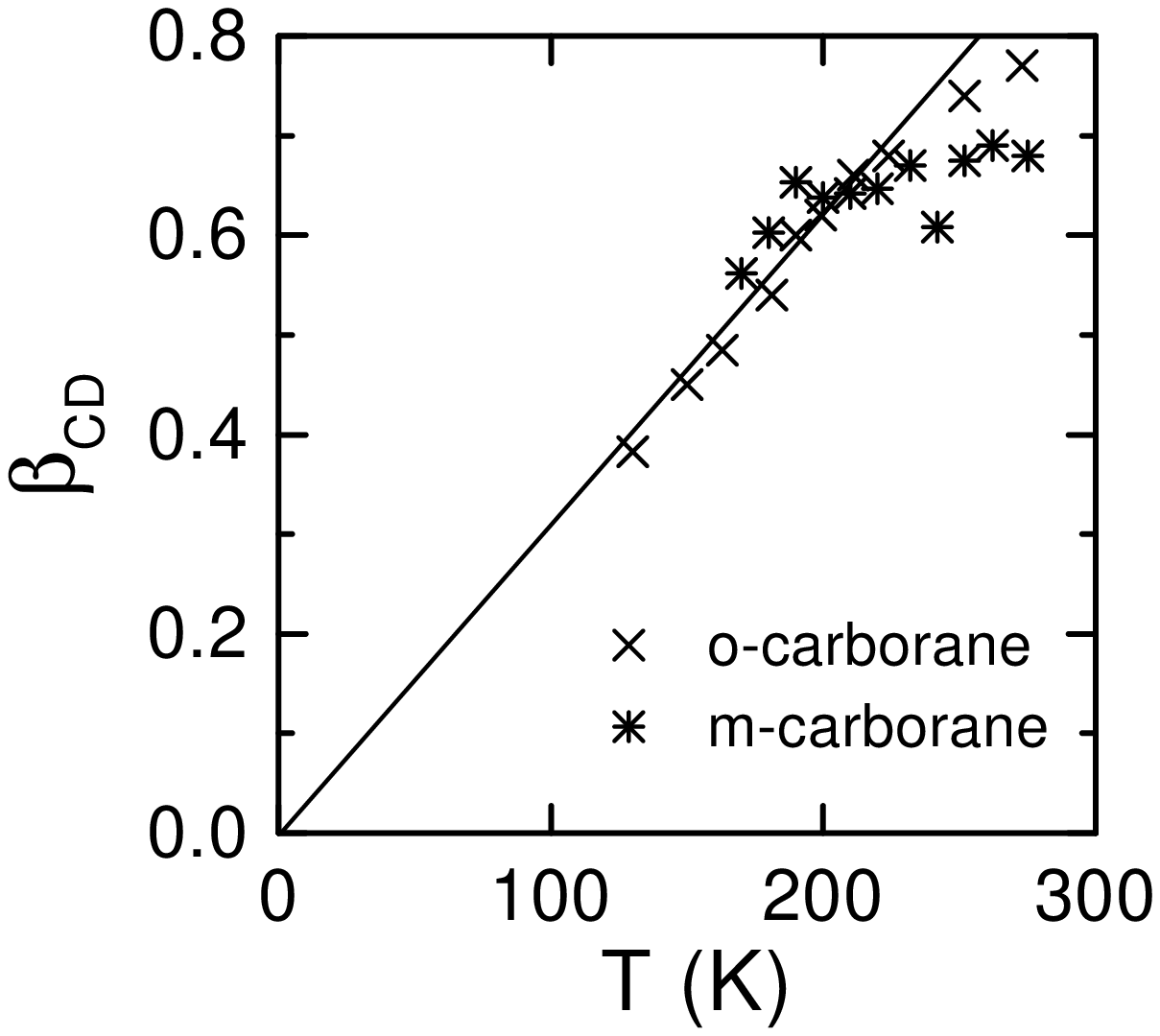}}
\caption{Temperature dependence of $\protect\beta _{CD}$ from fits of the $\protect\alpha
$-peak with the CD function for ortho- and meta-caborane. The solid line demonstrates a
possible extrapolation to $b(T=0)=0$.} \label{betapx}
\end{figure}

For the plastic crystals, the situation is different, as here the wing is missing and a
''divergent'' susceptibility would only be reached if $\beta (T)$ approached zero at a
finite temperature. Figure~\ref{betapx} shows the temperature dependence of $\beta _{CD}$
for o-CA and m-CA \cite {lunkenheimer96b,Brandunpubl}.
Indeed, $\beta _{CD}(T)$ weakly decreases with decreasing temperature. Due to the fact that these plastic crystals are
relatively ''strong'' glass-formers \cite{lunkenheimer96b,Brandunpubl}, $T_{VF}$\ is difficult to determine. It can be
expected to be located far below $T_{g}$ where it is not possible to obtain significant information from an
extrapolation of $\beta (T)$. The line in Figure~\ref{betapx} proposes a linear extrapolation of the low temperature
data to $\beta _{CD}(T=0{\,\mathrm K})=0$ but other extrapolations are also possible.

\section{Conclusions}

In summary, we have investigated the scaling of extremely broadband dielectric data using two different procedures. The
spectra on two glass-forming liquids and three plastic crystals have been analysed, extending the range of the abscissa
to higher values than in former analyses. For the glass-forming liquids, the approaches proposed by Nagel and coworkers
\cite{dixon90} and by Dendzik \textit{et al.} \cite{dendzik97} both lead to a satisfactory scaling of the loss data at
low frequencies. The extended frequency range available reveals a general failure of the Dendzik-scaling at the highest
frequencies investigated, while this effect is almost negligible for the Nagel-scaling. More broadband data on
glass-forming liquids are needed to check for the significance and universality of these deviations. If they were
confirmed, this may indicate that the wing is not so intimately related to the $\alpha $-relaxation as thought up to
now. Then it could be caused by processes, other than the $\alpha $-process (e.g. a $\beta $-relaxation
\cite{Leon99a,Leon99b}). But also a superposition of a very weak $\beta $-relaxation and the wing may be invoked to
explain the failing of the scaling at high frequencies \cite {kudlik95,leheny96,paluch}.

In addition, the broader frequency range of our data allowed for an extended check of the cross-over of the wing to
constant-loss behavior at a temperature near $T_{VF}$, claimed by Nagel and coworkers \cite {leheny97,menon95}. The
temperature dependence of the width parameter of the $\alpha $-peak gives no evidence for such a behaviour. For PC and
to a lesser extent also for glycerol, the temperature dependence of the wing exponent may be consistent with the
proposed approach of a divergent static susceptibility. However, the development of the power law prefactor for low
temperatures is still an open question. In our opinion the dielectric data reported in literature and even our extended
set of data do not provide a convincing experimental proof for a divergent susceptibility at low temperatures.

The data on the plastic crystals clearly cannot be scaled by the procedure proposed by Nagel and coworkers
\cite{brand99}. But using the Dendzik-scaling, all curves from the investigated plastic crystals fall onto one master
curve, different from the master curve of the glass-forming liquids. This master curve is identical with the master
curve for the Cole-Davidson function. This result corroborates the recent finding that plastic crystals without
intramolecular degrees of freedom do not show an excess wing but follow a CD-behaviour up to the highest frequencies
\cite{brand99}. In light of the missing excess wing, for plastic crystals a divergence of the static susceptibility at
finite temperatures seems unlikely, but cannot be fully excluded.

\section*{Acknowledgements}

We thank M. Fuchs and K.L. Ngai for valuable discussions. This work was supported by the
Deutsche Forschungsgemeinschaft, Grant-Nos. LO264/8-1 and LO264/9-1 and by the BMBF,
contract-No. 13N6917.

\end{document}